\documentclass[12pt]{iopart}

\newcommand{\sNN}[1]{$\sqrt{s_{NN}} = #1$ GeV}
\newcommand{\Lam}{$\Lambda $ }
\newcommand{\ALam}{$\bar{\Lambda} $ }
\newcommand{\Kaon}{$K^{0}_{S} $ }
\usepackage{graphicx}
\usepackage{epstopdf}

\begin{document}

\title[Strangeness production in jets from p+p \sNN{200} collisions]{Strangeness production in jets from p+p $\sqrt{s}  = 200$ GeV collisions}

\author{Anthony R. Timmins for the STAR Collaboration}

\address{Department of Physics and Astronomy, Wayne State University, 666 W. Hancock, Detroit, MI 48201, USA}
\ead{tone421@rcf.rhic.bnl.gov}

\begin{abstract}

Measurements of strangeness production in jets help illuminate the QCD mechanisms in fragmentation. Furthermore, they provide a crucial baseline for heavy-ion studies where modifications in jet chemistry have recently been observed by the STAR experiment. We present new results on strange particle production in jets from p+p $\sqrt{s}$ = 200 GeV collisions measured by the STAR experiment. The fragmentation functions of the \Lam, \ALam and \Kaon particles are obtained using various jet finding algorithms, and then compared to various models. Strange particle ratios in jets will be obtained and compared to values obtained from the inclusive spectra. Finally, we will show jets tagged with leading strange baryons and mesons, in order to investigate whether gluon or quark jets can be isolated in this way.

\end{abstract}

\section{Introduction}

We report measurements of strange particle fragmentation functions in p+p $\sqrt{s}  = 200$ collisions by the STAR experiment. These results provide additional tests to QCD inspired models of fragmentation such as PYTHIA \cite{PYTHIA} or the Modified Leading Log Approximation (MLLA) \cite{MLLA1,MLLA2,MLLA4,MLLA5}, beyond the usual comparisons to charged hadron fragmentation functions.  Furthermore, evidence for the modification of jet chemistry in heavy-ion collisions relative to p+p, has recently been observed by the STAR experiment for Au+Au \sNN{200} collisions via inclusive spectra measurements \cite{QMSpectra}. These measurements showed the $K/\pi$ ratio from Au+Au collisions at $p_T > 6 $ GeV/c is significantly higher than the respective value in p+p collisions. It is therefore essential we constrain theoretical descriptions of identified particle fragmentation functions, if the modifications in heavy-ion jet chemistry are to be further understood. Finally, since fragmentation functions characterise production from hard processes, comparisons to inclusive spectrum measurements may help elucidate the contributions from hard processes to those measurements as a function of $p_{T}$.

\begin{figure}[b]
\begin{center}
\includegraphics[width = 1\textwidth]{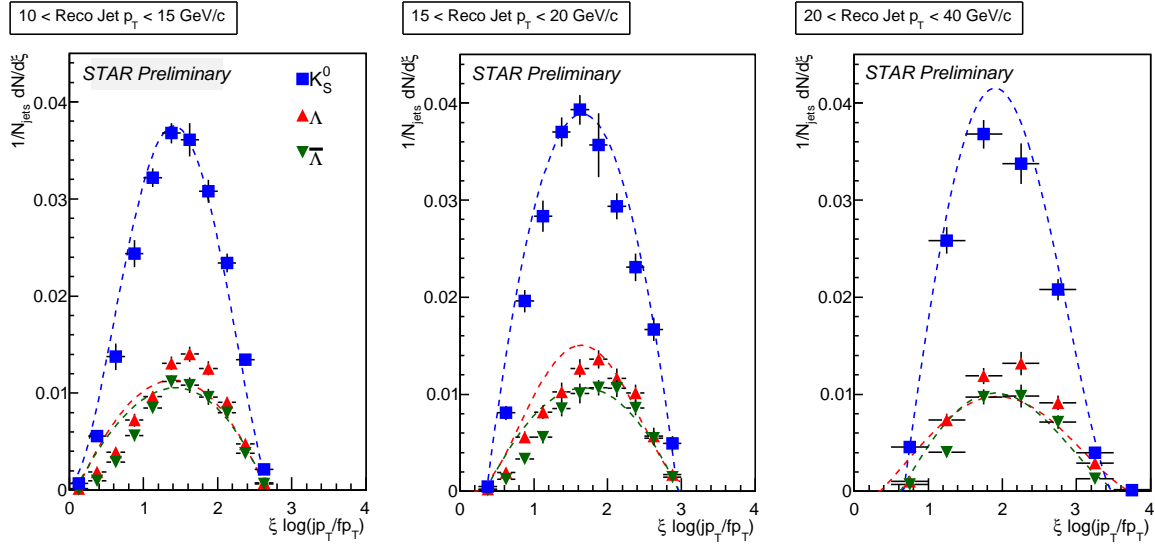}
\end{center}
\caption{Data points show strange particle fragmentation functions as a function of the reconstructed jet $p_{T}$ in p+p \sNN{200} collisions. The strange particle yields are not corrected for tracking efficiency. The symbols $jp_{T}$ and $fp_{T}$ are the reconstructed jet and fragment $p_{T}$ respectively. The curves show polynomial fits to PYTHIA 6.4 predictions. These predictions are obtained from PYTHIA events ran through the STAR detector simulator. The same applies to all other figures.} 
\label{fig1} 
 \end{figure}

\section{Analysis}

The data presented are from $\sim$ 8 million jet-patch triggered p+p \sNN{200} events recorded in 2006 by the STAR experiment. The jet-patch trigger is designed to select events with jets, and thus requires energies above 8 GeV to be deposited in an area $\Delta \eta \times \Delta \phi = 1 \times 1$ within the electromagnetic calorimeter \cite{STARCal}. The Time Projection Chamber (TPC) is used to detect the charged particles, while the barrel electromagnetic calorimeter, BEMC, is used for neutral particles. Three jet finders are employed from the fast-jet package \cite{fastjet}: $k_{t}$, anti-$k_{t}$ and SIS. We select recoil jets with respect to the triggered jets to measure fragmentation functions. This is because the triggered jets have a neutral energy bias. The jets are found within a resolution parameter, R $=0.4$ and a jet axis $\eta$ cut $< 0.6$ to ensure the jet area lies within the TPC/BEMC acceptance. The resolution parameter can be thought of as an effective jet radius in the $\eta,\phi$ plane. The minimum $p_{T}$ cut for a charged particle is 200 MeV/c, while the minimum calorimeter tower energy is 200 MeV. Towers which match charged particles are removed from the jet to avoid over counting of electrons and Minimum Ionising Particles (MIPs). %The sum of charged particle $p_T$ and tower energies is taken to be the reconstructed jet $p_{T}$.  
No corrections are applied to the jet energy scale which means the reconstructed jet $p_T$ is likely to be less than the actual jet $p_{T}$ due to TPC/BEMC detection inefficiencies, and missing neutral energy principally from undetected $n$ and $K^{0}_{L}$ particles. Studies are underway to determine a correction for this. Finally, in order to measure the fragmentation functions of the $\Lambda$, $\bar{\Lambda}$, and $K^{0}_{S}$ particles, V0s are identified within the jet area from TPC tracks. The invariant mass distributions are calculated to extract the yields of  these particles via their dominant decay channels: $\Lambda \rightarrow p+\pi^{-}$, $\bar{\Lambda} \rightarrow \bar{p}+\pi^{+}$, $K^{0}_{S} \rightarrow \pi^{+}+\pi^{-}$. In conjunction with a minimum $p_T$ cut of 1 GeV/c, a series of topological cuts are placed on the V0s to minimise the relative background contribution, and ensure the signal to background ratios are approximately the same for each particle. The residual background after all cuts is not yet removed from the fragment yields.

\section{Results}

Strange particle fragmentation functions as a function of the reconstructed jet $p_T$ are shown in figure \ref{fig1}. The fragmentation function is expressed in $\xi=log(jp_{T}/fp_{T})$, where $jp_{T}$ and $fp_{T}$ are the reconstructed jet and fragment $p_{T}$ respectively.  For a particular particle species, the integral of the $\xi$ function gives the mean number per jet over a chosen range in $\xi$. The uncertainties are statistical added in quadrature to the small differences obtained by comparing the three different jet finders. The measured V0 yields are not yet corrected for acceptance and TPC tracking inefficiencies. The PYTHIA predictions are obtained by processing PYTHIA events with STAR's detector simulation and reconstruction software. This ensures the PYTHIA events are subject to the same acceptance restrictions/detector inefficiencies as the real data. The jet/V0 finding is then run on the simulated output in the same way as the real data. All PYTHIA predictions shown in these proceedings are from version 6.4 with Tune A. We find that PYTHIA gives a reasonable description of the \Kaon data. JETSET (the jet production scheme in PYTHIA) has also been shown to describe $K^{\pm}$ fragmentation functions in $e^{+}+e^{-}$ collisions well, where the jet energies range $5 < E_{jet} < 46$ GeV \cite{FAnulli}. Furthermore, PYTHIA has been shown to describe charged hadron fragmentation functions in p+p collisions at the same center of mass energy  \cite{HelenQM}. On the other hand, we note that the PYTHIA description of \Lam and \ALam fragmentation functions are less satisfactory. Although the predictions appear to predict the correct yield over all $\xi$, they tend to over predict the yields at low $\xi$ and under predict the yields at intermediate $\xi$.

\begin{figure}[h]
\begin{center}
\includegraphics[width = 1\textwidth]{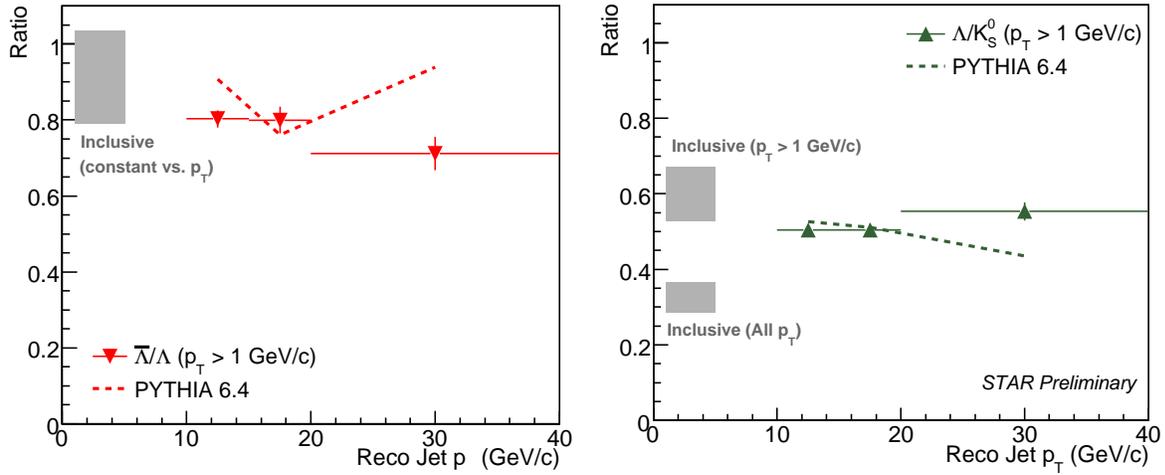}
\end{center}
\caption{ $\bar{\Lambda}/\Lambda$ and $\Lambda/K^{0}_{S}$ ratios in jets as a function of the reconstructed jet $p_{T}$. In each case, the minimum particle $p_{T}$ is 1 GeV/c. The uncertainties are statistical added in quadrature to the small differences obtained by comparing the three different jet finders. Inclusive refers to values obtained from inclusive spectra measurements \cite{STARpp}. The vertical size of shaded boxes reflects statistical and systematic uncertainties added in quadrature.}
\label{fig2} 
 \end{figure}

In figure \ref{fig2}, we show integrated $\bar{\Lambda}/\Lambda$ and $\Lambda/K^{0}_{S}$ ratios for $p_{T} > 1$ GeV/c as a function of reconstructed jet $p_T$. The analysis shows the $\bar{\Lambda}/\Lambda$ ratio is below 1 showing the baryon asymmetry in the colliding system (p+p) is transferred to jets over all jet energies measured. We also find the $\bar{\Lambda}/\Lambda$ ratio is consistent with the value obtained from inclusive spectra measurements \cite{STARpp}. PYTHIA also seems to reproduce the magnitude of the ratio. The jaggedness of the predictions is due statistical fluctuations which are not shown. Regarding the $\Lambda/K^{0}_{S}$ ratios in jets, these are higher than values from inclusive spectrum measurements over all $p_{T}$. The inclusive spectrum measurement will be dominated by production below 1 GeV/c. When we compare the ratios in jets to the inclusive spectrum measurement above 1 GeV/c, we find they are generally consistent. This may mean that $\Lambda$ and $K^{0}_{S}$ spectrum measurements with $p_{T} > 1$ GeV/c have a dominant contribution from hard processes i.e. jet production. We also find PYTHIA again reproduces the magnitude of the ratio.

\begin{figure}[h]
\begin{center}
\includegraphics[width = 0.6\textwidth]{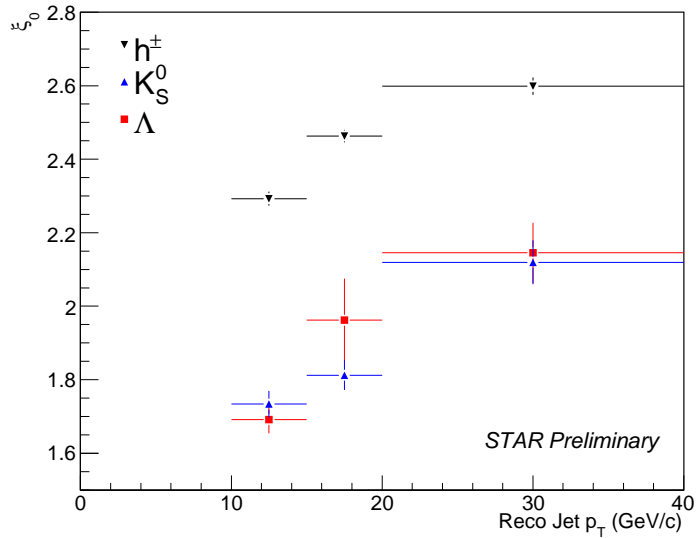}
\end{center}
\caption{Peak position of $\xi$ ($\xi_{0}$) for charged hadrons, \Lam and \Kaon particles as a function of reconstructed jet energy. The uncertainties are statistical added in quadrature to the small differences obtained by comparing the three different jet finders.  }
\label{fig3} 
 \end{figure}

In figure \ref{fig3}, we show the peak position of the $\xi$ distribution, $\xi_{0}$, for the various particles as a function of the reconstructed jet $p_{T}$. Since the $p_{T}$ dependent tracking inefficiency/acceptennce will distort the shape of the $\xi$ distribution, we correct for both of these effects. A Gaussian is fitted in the peak region to extract $\xi_{0}$. The fit range is chosen to avoid the region in $\xi$ effected by the minimum $p_T$ cut on the fragments. In the MLLA scheme, the peak position can be described by the following relation \cite{MLLA5};
\begin{equation}
\xi_{0} = Y + \sqrt{cY}-c
\label{equ:MLLA1}
\end{equation}
where $c$ is a constant and $Y$ is:
\begin{equation}
Y=log \left( \frac{E_{jet}sin(\theta_{0})}{Q_{eff}} \right)
\label{equ:MLLA1}
\end{equation}
where $\theta_{0}$ is the jet opening angle, and $Q_{eff}$ is the effective momentum where a parton ceases to branch in the parton shower. MLLA predictions for fragmentation functions of hadrons with mass $M_{0}$ typically assume $Q_{eff} \simeq M_{0}$ \cite{SapWeid}. Equation \ref {equ:MLLA1} then implies a scaling behaviour of $\xi_{0}$ with the mass of the hadron i.e.  $\xi_{0}$ should decrease for the higher mass hadrons. The value of $\xi_{0}$ should also increase with jet energy for a given hadron which is observed in figure \ref{fig3} for all particles. However, we find the mass scaling appears only to work to first order since although $\xi_{0}(h^{\pm}) > \xi_{0}(K^{0}_{S})$ and $\xi_{0}(\Lambda)$, $\xi_{0}(K^{0}_{S}) \sim \xi_{0}(\Lambda)$. A similar observation was made for $\pi,K,p$ fragmentation functions for similar jet energies in $e^{+}+e^{-}$ collisions \cite{FAnulli1}.

\begin{figure}[h]
\begin{center}
\includegraphics[width = 1\textwidth]{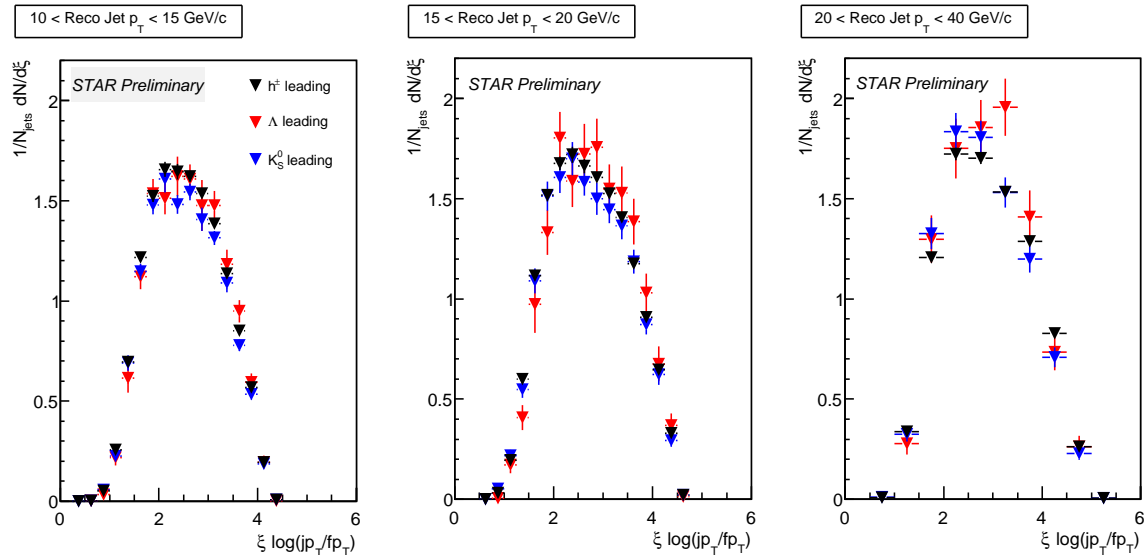}
\end{center}
\caption{Fragmentation functions of non-leading charged hadrons as a function of the reconstructed jet $p_{T}$ in p+p \sNN{200} collisions. The black points show data for jets where the leading particle is charged hadron, red corresponds to jets with leading \Lam particles, and the blue points correspond to jets with leading \Kaon particles.  The charged hadron yields are not corrected for tracking efficiency. $jp_{T}$ and $fp_{T}$ are the reconstructed jet and fragment $p_{T}$ respectively.  The uncertainties are statistical added in quadrature to the small differences obtained by comparing the three different jet finders.  } 
\label{fig4} 
 \end{figure}

Finally, in figure \ref{fig4} we show the fragmentation functions of \emph{non-leading} charged hadrons for the cases where the leading particle is a charged hadron, $\Lambda$, or $K^{0}_{S}$. We want to investigate whether tagging jets according to the species of the leading hadron, preferentially selects on gluon or quark jets. Both measured data \cite{DELPHJets} and theoretical predictions show hadron multiplicities are larger in gluon jets compared to quark jets: MLLA gives the ratio 9/4 for gluon/quark jet multiplicities. Thus, if tagging jets with particle A preferentially selected gluon jets, and tagging jets with particle B preferentially selected quark jets, we would expect the jets associated with particle A to have more non-leading charged hadrons compared to jets associated with particle B. However, in figure \ref{fig4} we observe at given jet energy, the charged hadron multiplicities are the same for each of the tagged jets. This might be expected for the highest jet $p_T$ bin where jet production is expected to be dominated by hard scattering of valance quarks, however at lower jet energies there should be mixture of quark and gluon jets \cite{GluonQuark}. Therefore, our observation suggests quark/gluon jets cannot be tagged in this way. Further studies are underway to confirm this.
%This is surprising since from baryon number conservation, one may naively expect quark jets to preferentially produce leading baryons, while gluon jets leading mesons.

\section{Summary}

In summary, we have shown measurements of strange particle fragmentation functions. We have found that PYTHIA describes the $K^{0}_{S}$  fragmentation functions well, however we observe some deviations for the $\Lambda$ and $\bar{\Lambda}$ data. We have found that $\bar{\Lambda}/\Lambda$ ratios ($p_{T} > 1$ GeV/c) in jets are consistent with values obtained from inclusive spectra. ${\Lambda}/K^{0}_{S}$ ratios  ($p_{T} > 1$ GeV/c) in jets were found to be similar to values obtained from inclusive spectra in the same $p_T$ range. This may suggest that ${\Lambda}$ and $K^{0}_{S}$ production above 1 GeV/c is dominated by hard processes in p+p \sNN{200} collisions. We investigated a mass scaling of the peak position of $\xi$ (inferred from MLLA), and found although $\xi_{0}(h^{\pm}) > \xi_{0}(K^{0}_{S})$ and $\xi_{0}(\Lambda)$, $\xi_{0}(K^{0}_{S}) \sim \xi_{0}(\Lambda)$. Finally, we have found tagging jets with leading strange baryons/mesons may not preferentially select on quark/gluon jets.

\end{document}